\input{aipcheck}

\documentclass[final
    ,final            
  ]
  {aipproc}

\layoutstyle{6x9}

\def\littleprime{\ifmmode{\scriptscriptstyle \prime }
    \else{\hbox{$\scriptscriptstyle \prime$ }}\fi}
\def\littlecirc{\ifmmode{\scriptscriptstyle \circ }
    \else{\hbox{$\scriptscriptstyle \circ $ }}\fi}
\def\littless{\ifmmode{\scriptscriptstyle s }
    \else{\hbox{$\scriptscriptstyle s $ }}\fi}
\def\arcsec{\raise .9ex \hbox{\littleprime\hskip-3pt\littleprime}}
\def\arcmin{\raise .9ex \hbox{\littleprime}}
\def\degree{\raise .9ex \hbox{\littlecirc}}
\def\arcsecpoint{\hbox to 1pt{}\rlap{\arcsec}.\hbox to 2pt{}}
\def\arcminpoint{\hbox to 1pt{}\rlap{\arcmin}.\hbox to 2pt{}}
\def\degreepoint{\hbox to 1pt{}\rlap{\degree}.\hbox to 2pt{}}
\def\gtapr {\lower .1ex\hbox{\rlap{\raise .6ex\hbox{\hskip .3ex
        {\ifmmode{\scriptscriptstyle >}\else
                {$\scriptscriptstyle >$}\fi}}}
        \kern -.4ex{\ifmmode{\scriptscriptstyle \sim}\else
                {$\scriptscriptstyle\sim$}\fi}}}
\def\ltapr {\lower .1ex\hbox{\rlap{\raise .6ex\hbox{\hskip .3ex
        {\ifmmode{\scriptscriptstyle <}\else    
                {$\scriptscriptstyle <$}\fi}}}
        \kern -.4ex{\ifmmode{\scriptscriptstyle \sim}\else
                {$\scriptscriptstyle\sim$}\fi}}}

\begin{document}

\title{Improved constraints on the cosmological parameters using the VLA FIRST survey}


\author{E. Xanthopoulos}{
  address={University of California Davis, Department of Physics, Davis, CA 95616}
  ,altaddress={IGPP/Lawrence Livermore National Laboratory, Livermore, CA 94550 }
}

\author{R. H. Becker  }{
  address={University of California Davis, Department of Physics, Davis, CA 95616}
  ,altaddress={IGPP/Lawrence Livermore National Laboratory, Livermore, CA 94550 }
}

\author{W. H. deVries  }{
   address={University of California Davis, Department of Physics, Davis, CA 95616}
  ,altaddress={IGPP/Lawrence Livermore National Laboratory, Livermore, CA 94550 }
}

\author{R. L. White   }{
  address={Space Telescope Science Institute, 3700 San Martin Drive, Baltimore, MD 21218}
}

\begin{abstract}
Using the final version of the VLA FIRST survey (Becker et al.
2003), the optical Sloan Digital Sky Survey Data Release 3 (DR3)
quasar list and a series of carefully selected criteria, we have
defined the largest homogeneous population of double-lobed
sources. With the precise sample definition, the increased depth
and sensitivity of the survey data, the large size of the dataset,
and our self-consistent method of analysis, which addresses many
of the problems associated with previous work in the area, we are
able to: a) explore the correlations between the intrinsic
properties of the double-lobed radio sources (the results are also
confirmed by a non-parametric analysis) and study their evolution,
b) place more interesting and tighter constraints on the
cosmological parameters, distinguish among the different cosmology
models, and determine the impact of the angular size-redshift
studies in cosmology, c) further our understanding of the behavior
of the intergalactic medium (IGM) density as a function of
redshift and shed more light to the quasar-radio galaxy
unification issue.

\end{abstract}

\maketitle


\section{Introduction}

The angular size-redshift (theta-z) relation for a cosmological
population of standard rods is a powerful probe of the large-scale
geometry of the Universe. However, previous attempts to measure
the cosmological parameters from the angular size-redshift
relation of double radio-lobed radio sources have been marred by a
variety of selection effects, destroying the integrity of the data
sets, and by inconsistencies in the analysis, undermining the
results and leading to data consistent with a static Euclidean
Universe rather than with standard Friedmann models
(e.g. Wardle \& Miley 1974, Barthel \& Miley 1988, Nilsson et al. 1993).
Interpretation of this observation has caused disagreements among
various authors.

\begin{figure}[h]
\caption{The angular size - redshift relation for deprojected rods
of length 100 h$_{0}^{-1}$ kpc for different cosmologies. The
choice of the cosmological parameters for the three Friedmann
models are listed on the figure. The curve for a static Euclidean
universe is shown for comparison. In practice, the curves actually
define upper limits to the observed angular sizes, since
projection effects will scatter the observed sizes downward. Note
the presence of the minimum near z$\sim$1.5.}\label{FRWe}
\includegraphics[height=.35\textheight]{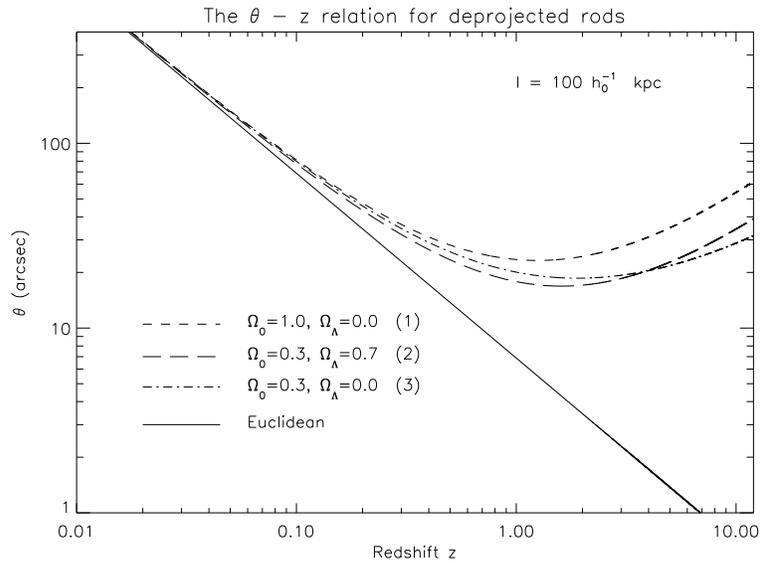}
\end{figure}


\section{The sample and selection criteria}
The VLA FIRST survey (Becker et al. 1995) has mapped $\sim$9000
deg$^{2}$ of the sky at 1.4 GHz to a sensitivity of $\sim$1 mJy
with 5\arcsecpoint4 FWHM Gaussian beam and has cataloged 111,115
sources with subarcsecond positional accuracy. We cross-correlated
this survey with a sample of 51,516 quasars compiled from the
optical Sloan Digital Sky Survey Data Release 3 (DR3) quasar list
and the list of quasars from the 2dF survey. By using a lower
detection limit of 3$\sigma$ for the Poisson probability
distribution of finding companions of the quasars within an area
of 5$\arcmin$ radius from the quasar position, we defined an
initial sample of double-lobed radio sources with their cores
centered on the quasar positions. We then followed a series of
carefully selected criteria, and specifically we: a) included only
radio sources, FRIIs, with symmetric and collinear triple
structure (core  + two lobes), thereby minimizing asymmetrical
effects that might distort the apparent angular size, such as
relative motion with respect to the IGM, b) further restricted the
sample to sources with redshift z$>$0.3 beyond which quasars begin
to dominate and carry more info about the cosmology. This way we
bypass the problem of different mean orientations or power-size
correlations which produce non- cosmological effects in the
theta-z plane. Our final clean sample consists of 389 FRIIs, which
is the largest homogeneous population of double-lobed sources to
date.

\section{The angular size - redshift relationship}

Fig.~\ref{FRWe} illustrates the $\theta - z$ relationship for
deprojected rods ($\phi$ = 90$^\circ$) with an intrinsic size of
l=100 h$_{0}^{-1}$ kpc for three Friedmann cosmologies: a) an
Einstein-de Sitter universe, b) a flat universe, and 3) a
nonclosed, matter-dominated universe. The particular values of
$\Omega_{0}$ and $\Omega_{\Lambda}$ are listed in the figure. The
curve for a static Euclidean universe, with $\theta \propto
z^{-1}$ is also shown for comparison. The location of the minimum
in the angular size (typically between z=1 and z=2) depends on
$\Omega_{0}$ and $\Omega_{\Lambda}$.

\begin{figure}[h]
\begin{minipage}[t]{7.4cm}
\includegraphics[width=1\textwidth]{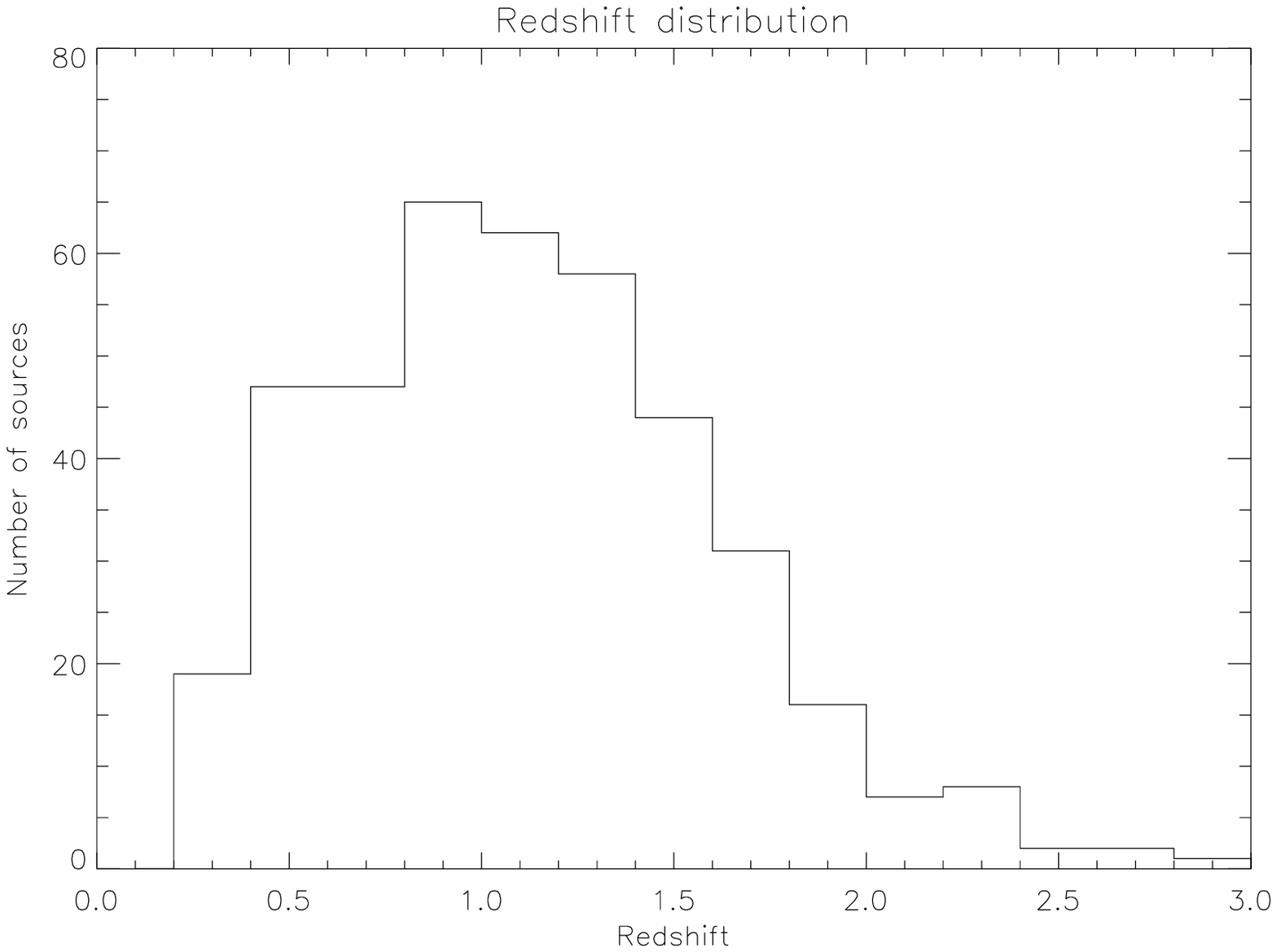}
\end{minipage}
\hfill
\begin{minipage}[t]{7.4cm}
\includegraphics[width=1\textwidth]{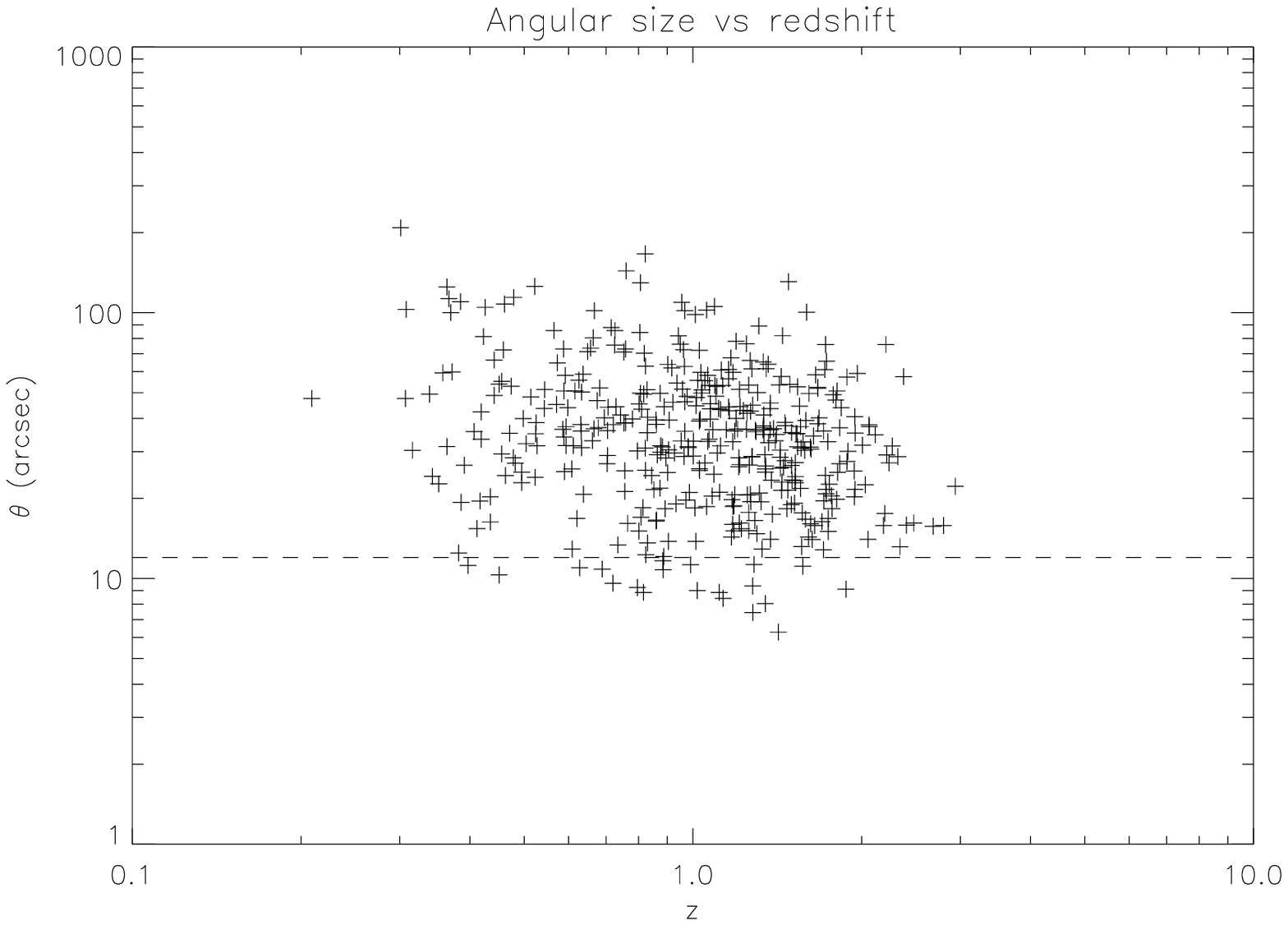}
\caption {(Left:) The distribution of redshift in our sample. The
sample includes sources out to a redshift of 2.94 significantly
higher than the redshifts at which the minima in the theta-z
curves typically occur for different Friedmann models, and in
contrast to previous work which included samples that contained
significant numbers of sources with z$<$1 where roughly Euclidean
behavior is expected. The median redshift of our sample is 1.074
while that of the FIRST survey is ~1, evidence that indicates that
the two populations have similar distributions, and therefore our
sample selected using the optical SDSS survey is fairly
representative of double-lobed radio sources as a whole. (Right:)
Scatter plot of the peak-to-peak angular sizes vs redshift. The
dashed line represents the effective resolution limit at
12\arcsec, below which accurate morphological classifications
could not be determined. The errors in the measured values of the
angular size are typically $\sim$1\arcsec, far less than the
scatter in the angular sizes at any redshift.} \label{thetaz}
\end{minipage}
\hfill
\end{figure}

\noindent Fig.~\ref{thetaz} shows a scatter plot of the $\theta -
z$ data. The errors in the measured values of $\theta$ are
typically $\sim$1\arcsec, far less that the scatter in the angular
sizes at any redshift. With 5\arcsecpoint4 FWHM beam, the FIRST
survey can detect extended structure down to 2\arcsec. However,
due to the survey resolution limit, uncertainties in the quasar
optical positions, variations in the morphologies of double-lobed
sources, and inspection of numerous FIRST radio maps, we also
introduce an effective cutoff in the data of 12\arcsec, shown in
Fig.~\ref{thetaz}, following Buchalter et al. (1998), below which
an accurate morphological classification could not be assigned
with certainty. Such a cutoff ensures that we do not include the
so called core-jet, diffuse, cometary and other type of extended
radio sources that may be mistaken for double-lobed objects at low
resolution.

\section{Standard and Nonstandard cosmological models}

\begin{figure}[h]
\begin{minipage}[t]{7.4cm}
\includegraphics[width=1\textwidth]{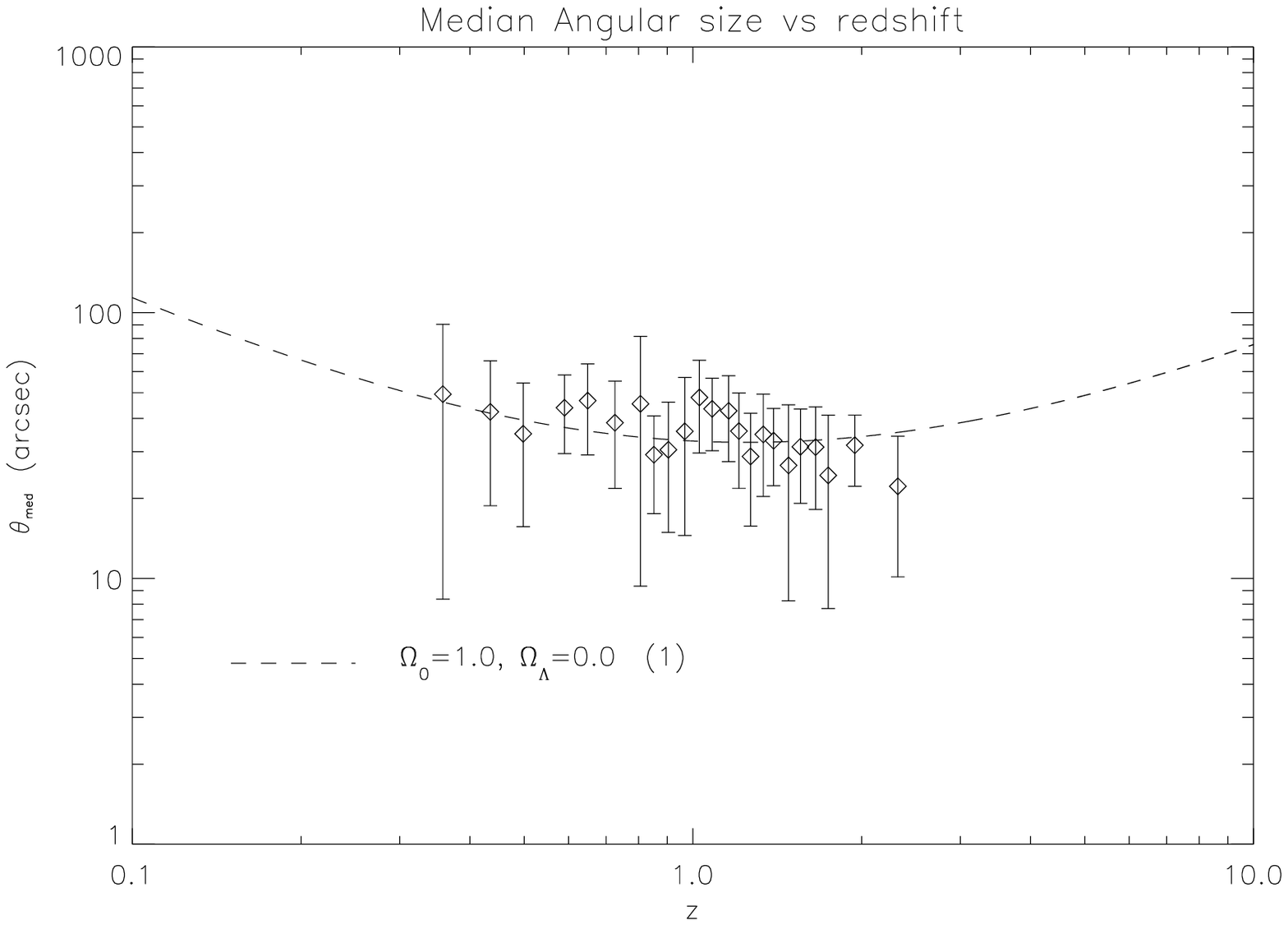}
\end{minipage}
\hfill
\begin{minipage}[t]{7.4cm}
\includegraphics[width=1\textwidth]{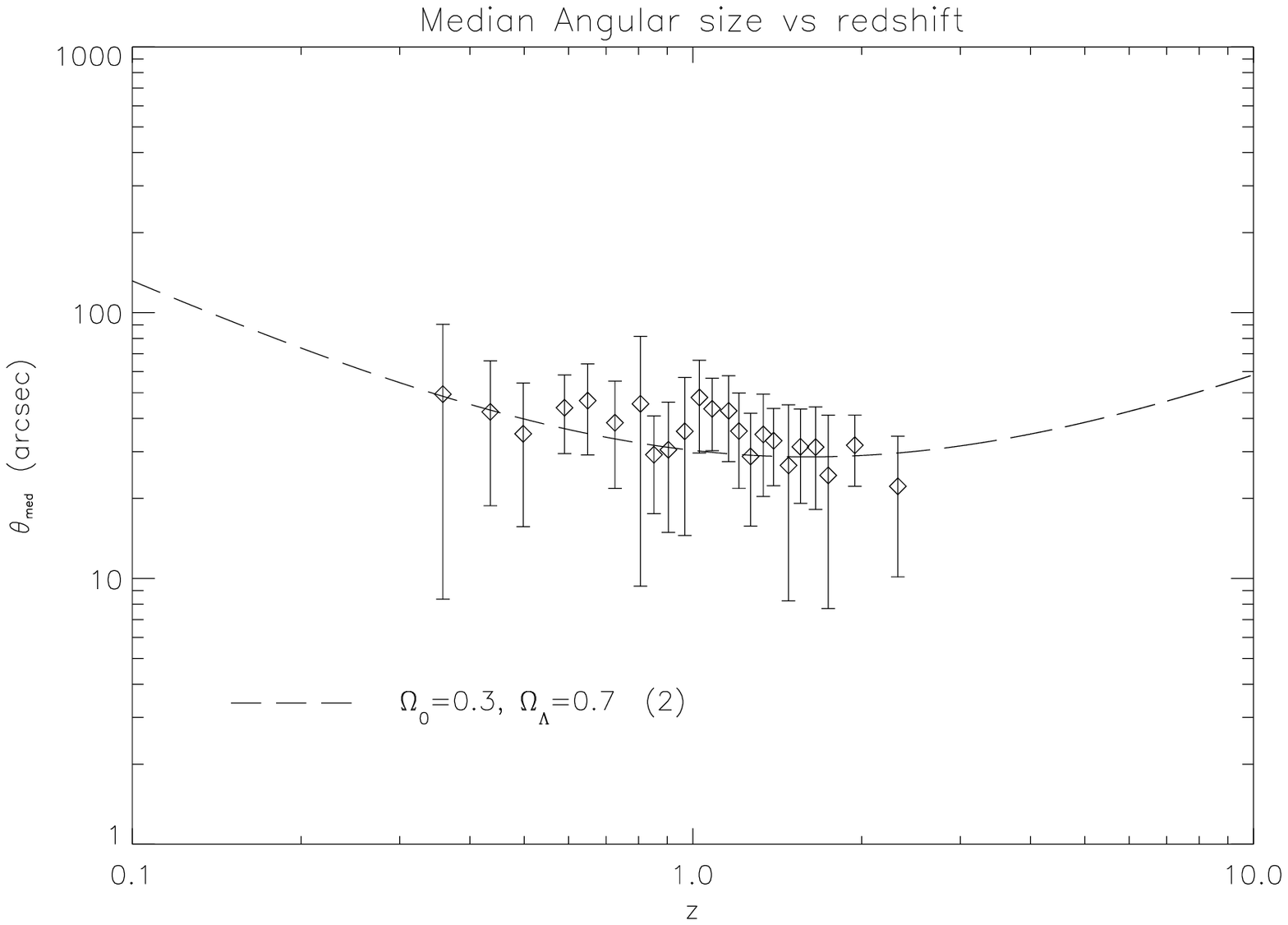}
\caption{Median angular sizes, binned in redshift (to compensate
for projection effects), with roughly equal numbers per bin
($\sim$square-root of N the number of sources in the sample). The
Friedmann curves, corresponding to models 1 and 2 from
Fig.~\ref{FRWe}, have been scaled to provide a rough visual fit.
The error bars represent the median absolute deviation in each
bin.} \label{model12}
\end{minipage}
\hfill
\end{figure}

\begin{figure}[h]
\begin{minipage}[t]{7.4cm}
\includegraphics[width=1\textwidth]{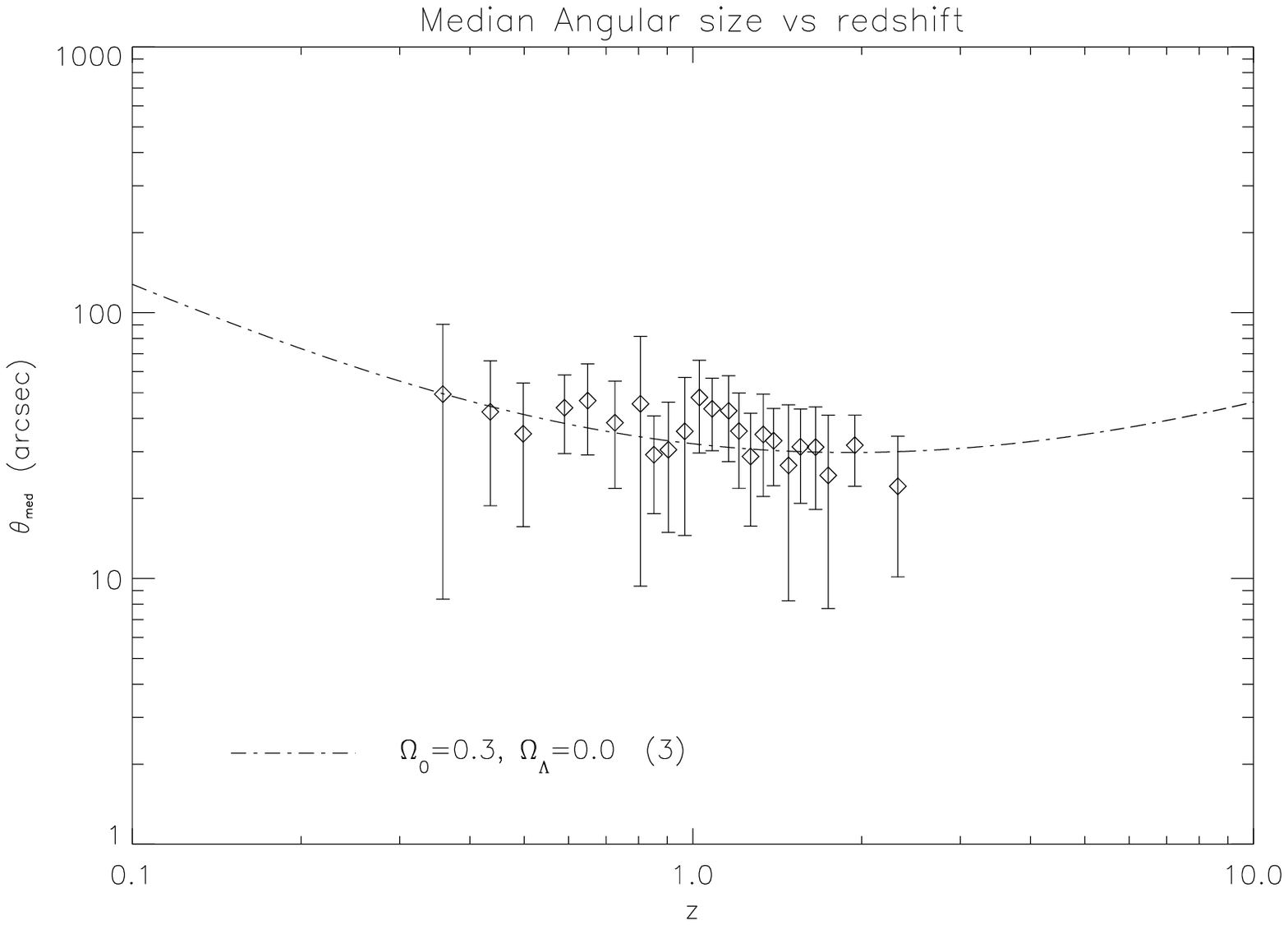}
\end{minipage}
\hfill
\begin{minipage}[t]{7.4cm}
\includegraphics[width=1\textwidth]{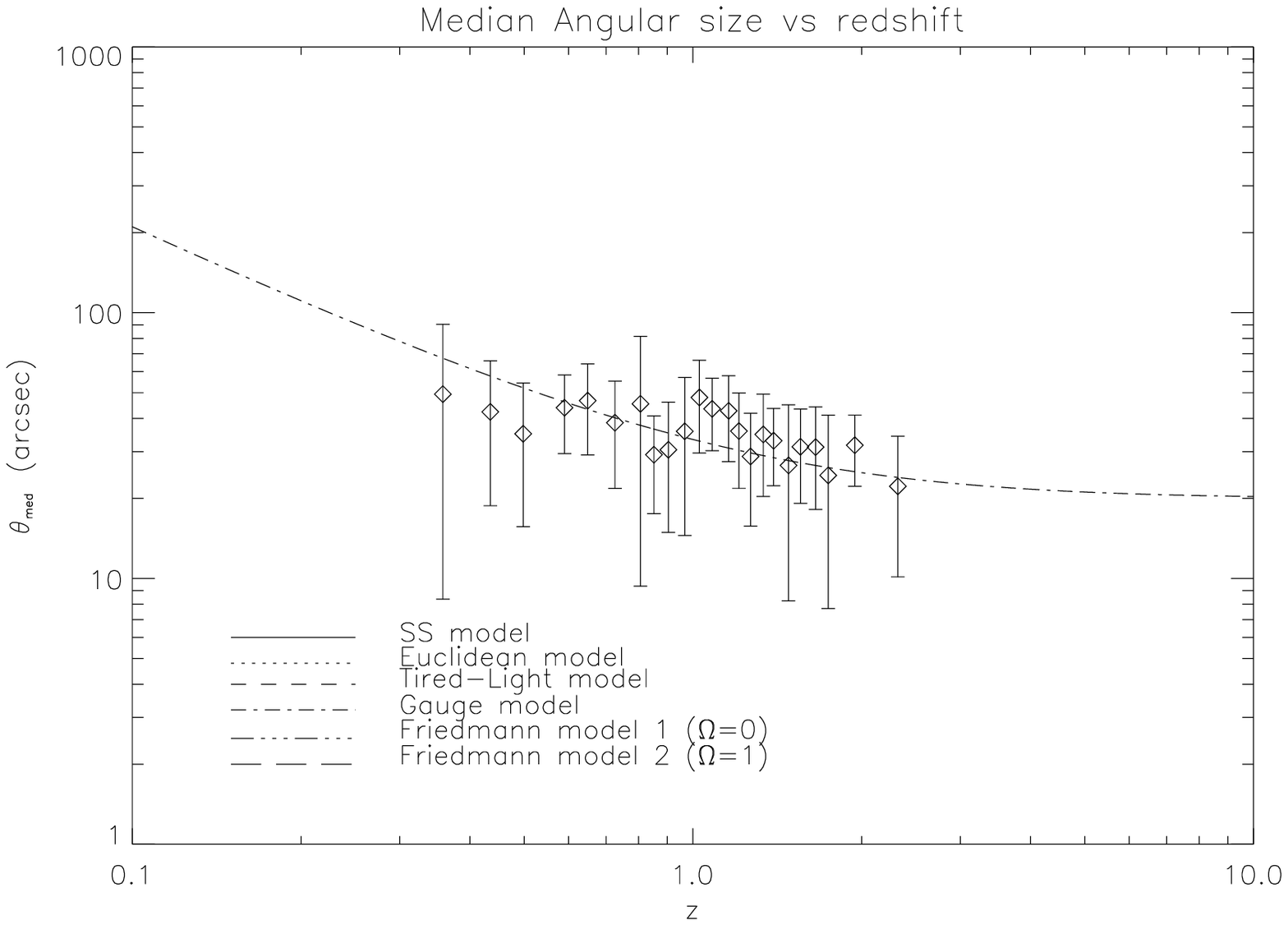}
\caption {(Left:) Median angular sizes, binned in redshift (to
compensate for projection effects), with roughly equal numbers per
bin ($\sim$square-root of N the number of sources in the sample).
The Friedmann curve, corresponding to model 3 from
Fig.~\ref{FRWe}, has been scaled to provide a rough visual fit.
The error bars represent the median absolute deviation in each
bin. (Right:) The Gauge curve has been scaled to provide a
qualitative reference. In the Gauge model, when z tends to
infinity the angular size tends to a constant similar to the
Friedmann model with qo=0 but more rapidly.} \label{model34}
\end{minipage}
\hfill
\end{figure}

For graphical purposes, we bin the data in redshift, in equal
numbers per bin, and calculate the mean and median values of the
theta, together we standard errors of the mean values and median
absolute deviations for each bin. The results are shown in
Fig.~\ref{model12} and Fig.~\ref{model34}, for the median values
of $\theta$, along with the curves from Fig.~\ref{FRWe}, whose
amplitudes (corresponding to the median intrinsic sizes), have
been scaled to provide a rough visual fit. Apart from the
Friedmann models we have also experimented with the Steady State
and two nonstandard cosmology models, the Tired-Light and the
Gauge models, the latter shown in Fig.~\ref{model34}. The most
striking feature of the data is that, regardless of the binning
details, the observed data seem to be more consistent with
Friedmann models than with a Euclidean model. The Friedmann curves
shown are not the best-fit results but are merely intended for
qualitative reference.

\section{Discussion and Future work}

Before attempting to find the best-fit cosmological parameters,
and determine whether we can distinguish with high significance
between the different models with the present sample, we need to
explore the relationships between the intrinsic properties (P
intrinsic power, l projected linear size and redshift z) of the
sources in our sample using both parametric and non-parametric
analysis. Such correlations have important implications in
determining the parameters. Therefore our next steps are to: a)
Optimize the analysis by defining and exploring a parameter space
and use an iterative procedure that will lead to the best model
fit parameters, b) Apply a chi-square goodness-to-fit statistic to
determine the best fit values for the free parameters in each
model and, c) Examine the values of the Hubble constant implied by
the data.


\begin{theacknowledgments}
EXs work was performed under the auspices of the U.S. Department
of Energy, National Nuclear Security Administration by the
University of California, Lawrence Livermore National Laboratory
under contract No. W-7405-Eng-48 and she also acknowledges support
from the National Science Foundation (grant AST 00-98355).
\end{theacknowledgments}






\end{document}